\documentclass[aps,10pt,pra,twocolumn,groupedaddress, floatfix, superscriptaddress,showpacs, showkeys,amsfonts]{revtex4-2}
\usepackage{amsfonts}
\usepackage{amsmath}
\usepackage{amssymb, setspace}
\usepackage{graphics,graphicx}
\graphicspath{{./}}
\usepackage{epsf}
\usepackage{subfigure}
\usepackage[%
  colorlinks=true,
  urlcolor=blue,
  linkcolor=blue,
  citecolor=blue
]{hyperref}
\usepackage[all]{hypcap}
\usepackage{color}
\usepackage[utf8]{inputenc}
\usepackage{dsfont}
\usepackage{enumitem}
\usepackage[utf8]{inputenc}

\usepackage{comment}
\usepackage{orcidlink}

\usepackage{natbib}
\usepackage{braket}
\usepackage{physics}

\newcommand{\be}{\begin{equation}}
\newcommand{\ee}{\end{equation}}
\newcommand{\bea}{\begin{eqnarray}}
\newcommand{\eea}{\end{eqnarray}}
\newcommand{\eq}[1]{Eq.~\eqref{#1}}

\newcommand{\fig}[1]{Fig.~\ref{#1}}

\newcommand{\bem}{\begin{multline}}
\newcommand{\eem}{\end{multline}}

\newcommand\identity{1\kern-0.25em\text{l}}

\newcommand{\superket}[1]{\left|\!\left.#1\right>\!\right>}
\newcommand{\superbra}[1]{\left<\!\left<#1\right.\!\right|}
\newcommand{\superketbra}[2]{\left|\!\left.#1\!\left>\!\left>\!\right<\!\right<\!#2\right.\!\right|}

\newcommand{\apporsm}[1]{Appendix~\ref{#1}}

\begin{document}
\title{Randomized Benchmarking Protocol for Dynamic Circuits}

\author{Liran Shirizly\orcidlink{0009-0002-4597-0126}}
\email{liran.shirizly@ibm.com}
\affiliation{IBM Quantum, IBM Research - Israel, Haifa University Campus, Mount Carmel, Haifa 31905, Israel}
\author{Luke~C.~G.~Govia\orcidlink{0000-0002-3263-648X}}
\affiliation{IBM Quantum, IBM Almaden Research Center, San Jose, CA 95120, USA}
\author{David C. McKay\orcidlink{0009-0003-3237-3071}}
\affiliation{IBM Quantum, IBM TJ Watson Research Center, Yorktown Heights, NY 10598, USA}

\begin{abstract}
    Dynamic circuit operations -- measurements with feedforward --  are important components for future quantum computing efforts, but lag behind gates in the availability of characterization methods. Here we introduce a series of dynamic circuit benchmarking routines based on interleaving dynamic circuit operation blocks $\mathcal{F}$ in one-qubit randomized benchmarking sequences of data qubits. $\mathcal{F}$ spans between the set of data qubits and a measurement qubit and may include feedforward operations based on the measurement. We identify six candidate operation blocks, such as preparing the measured qubit in $|0\rangle$ and performing a $Z$-Pauli on the data qubit conditioned on a measurement of `1'. Importantly, these blocks provide a methodology to accumulate readout assignment errors in a long circuit sequence. We also show the importance of dynamic-decoupling in reducing ZZ crosstalk and measurement-induced phase errors during dynamic circuit blocks. When measured on an IBM Eagle device with appropriate dynamical decoupling, the results are consistent with measurement assignment error and the decoherence of the data qubit as the leading error sources.
\end{abstract}

\maketitle

\section{Introduction}

A major milestone of quantum computing is to demonstrate fault-tolerant operations using a quantum error correcting code. While there have been several promising code demonstrations~\cite{Sundaresan:2023,ryananderson2024high,zhao:2022,google:2023,krinner:2022}, there has been no definitive demonstration of full fault-tolerant operation. This indicates that the various system components need further development. To validate this development, effective methods for benchmarking these components will be required. Among the elements required are one- and two-qubit gates, mid-circuit measurement, qubit reset, and feedforward operations. Given the importance of gates to all types of quantum algorithms, many different techniques exist to benchmark gates. In particular, gates are typically benchmarked with a variety of randomized benchmarking (RB) protocols~\cite{Helsen20,Knill08,Magesan11,Magesan12,Magesan12b,Gambetta12,Wallman15,Wallman16,Cross16,Wood18,McKay19,Helsen19,Proctor19,Erhard19,McKay20,Morvan21}. These run the gamut from looking at individual gates~\cite{Magesan12}, to full system holistic benchmarks~\cite{proctor:2022,hines:2023,mckay:2023}.  In its simplest form, a Clifford RB protocol follows these steps:
\begin{itemize}
\item Construct a random sequence of Clifford gates of depth $l$ with the inverse Clifford appended.
\item Transpile the Clifford gates into the basis gates of the system.
\item Run the circuit on hardware and measure the probability of returning to the starting state $P(0)$.
\item Fit $P(0)$ vs $l$ averaged over different random circuits to extract an average gate fidelity. 
\end{itemize}

Measurements, and specifically mid-circuit measurements (MCM), i.e.~measurements that occur in the middle of a circuit and have gate operations before and after, are less amenable to RB protocols because the measurement operation is not a Clifford (or even unitary) operation. However, there have been several extensions that attempt to capture the main important features of measurement, such as MCM RB~\cite{Govia_2023} and MCM Pauli learning~\cite{hines:2024,gupta:2023,zhang:2024,beale:2023}. Alternatively, full tomography can be applied~\cite{blumoff:2016,rudinger:2022,pereira:2022,Pereira_2023}. Crucially, unlike typical gate errors, the assignment and quantum non-demolition (QND) errors in measurement do not accumulate in the terminal measurement signal, which means that long sequences of measurements do not give an obvious advantage in measuring small errors without post-processing on the outcomes of the MCM's.\\

There is a powerful extension to a MCM - the feedforward operation (FF) - which together are described as a `dynamic circuit'. In a dynamic circuit, the MCM's outcomes are written to classical data registers, and then certain gate operations in the circuit are conditioned on the outcome of a classical computation applied to the classical data registers. These types of operations are important components for QEC (with some early demonstrations, e.g. Ref~\cite{riste:2013}), but can also be used in near-term algorithms for short-depth state preparation~\cite{steffen:2013,smith:2024,kang:2024,baumer:2023,Malz_2024,baumer2024measurementbased}, and can be used to combine processors~\cite{vazquez:2024} . Despite their interest, there are few methods to benchmark these dynamic operations. Therefore, here we show how dynamic circuits can be characterized using RB as an extension to our earlier MCM RB work~\cite{Govia_2023}. In particular, we again consider interleaving the MCM part of the dynamic circuit (`measured qubit') with 1Q RB on a `data qubit'. However, this time we perform feedforward operations on the data qubit based on the outcome of the measurement. \\

Critically, this allows the assignment error of the measurement to feed into the 1Q RB error on the data qubit and therefore accumulate with the number of dynamic circuit operations. These protocols also highlight other important errors in these operations, such as $T_1/T_2$ error due to finite measurement time (similar to MCM RB), and the delay time of the FF operation. The latter is mainly due to the finite time required to perform the classical compute of the FF decision~\cite{vazquez:2024}. Measurement-induced phase errors on data qubits and static ZZ crosstalk are another important error consideration for superconducting qubits, which we show can be greatly decreased using appropriate dynamical decoupling sequences. \\

Our paper is organized as follows. In \S~\ref{sect:protocol} we discuss the protocol in detail. In \S~\ref{sect:errors} we present simulations and theory of the circuits with simple error models to highlight the main error terms that are measured. Finally, in \S~\ref{sect:exp} we show experimental data running the protocol on an IBM device.

\section{The protocol \label{sect:protocol}}

\begin{figure}
\centering
\includegraphics[width=0.49\textwidth]{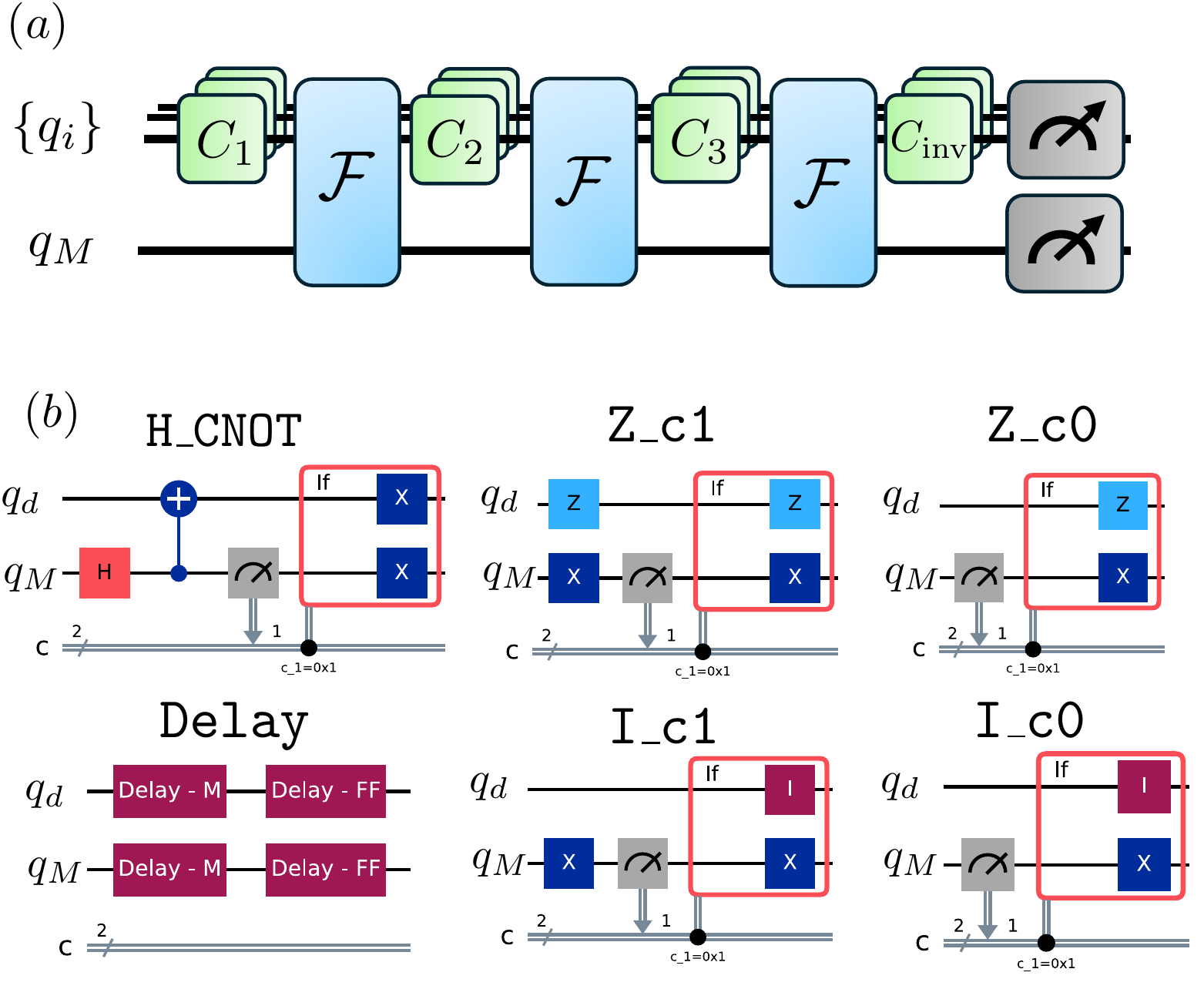}
\caption{\label{fig:ff_protocol} (a) Example of the protocol with $l=3$ sequence length. Sequences of one-qubit Clifford gates are applied on data qubits simultaneously, interleaved with dynamic circuit blocks $\mathcal{F}$ which expand the space to include a measurement qubit. Each $\mathcal{F}$ is ideally equivalent to the identity operation. (b) The specific dynamic circuit blocks $\mathcal{F}$ that we use in this work, as described in the main text.
}
\end{figure}

Similar to the MCM RB protocol described in Ref.~\cite{Govia_2023}, our dynamic circuit benchmarking protocol is based on the idea of interleaving an identity operation in a one-qubit RB sequence. Each protocol follows these steps:
\begin{enumerate}
    \item Construct $m$ random 1Q RB sequences of length $l$, $\left\{C_j\right\}$ ($j=\{0,\ldots,l\}$), on a set of data qubits $\left\{q_i\right\}$.
    \item After every $k$ Clifford gates in the sequence interleave a dynamic circuit block $\left\{ \mathcal{F} \right\}$ which is ideally an identity operation on the data qubits. This block acts on all the data qubits and an additional measurement qubit $q_M$.
    \item Measure the ground state probability of each of the data qubits versus $l$ and average over random 1Q RB sequences. 
    \item Fit to an exponential decay $P_{i}(0)=A_i\alpha_{i,\mathcal{F}}^{l/k}+B_i$ where $l/k$ is the number of dynamic circuit blocks.
    \item Ascribe an average gate error for each block on each data qubit $i$ in the standard way $\varepsilon_{i,\mathcal{F}} = \frac{1}{2}\left(1-\alpha_{i,\mathcal{F}}\right)$.
\end{enumerate}
The protocol is illustrated in \fig{fig:ff_protocol}, along with several proposed dynamic circuit blocks $\mathcal{F}$. These blocks contain gates, measurements and feedforward operations such that their ideal operation is the identity, thus preserving the separation between the data qubits and measurement qubit subspace. We assume that the error is predominantly from $\mathcal{F}$ so that the average block fidelity on the data qubit can be obtained exclusively from the interleaved curve, i.e., without a reference curve. \\

Unlike in Ref.~\cite{Govia_2023}, the addition of the feedforward operation allows for non-trivial gate operations that can be canceled by the measurement and feedforward. We separate these circuits into two cases. In the first case, $\mathcal{F}$ includes two-qubit gates between the data qubit(s) and the measurement qubit. However, this requires connectivity between this pair of qubits. In the second case, there are no two-qubit gates which removes the connectivity constraint. The first case is a good model for a data-ancilla pair in an error correcting code. The second case is a good pure test of the feedforward operation and mixing of assignment error into the one-qubit gate error. In the case where $\mathcal{F}$ is just a measurement or measurement delay this replicates the MCM RB procedure of Ref.~\cite{Govia_2023}.

Here we consider several key circuits for $\mathcal{F}$ which are also shown in \fig{fig:ff_protocol}. 
\begin{itemize}
    \item \texttt{H\_CNOT}: A Hadamard ($H$) gate is applied to $q_M$ followed by a CNOT between $q_M$ and the data qubit $q_d$ where $q_M$ is the control. This is followed by a measurement on $q_M$; if the measurement is `0' then the operation of the CNOT on $q_d$ was the identity and if the measurement is `1' the operation was an $X$ gate. Therefore, conditioned on classical outcome `1', an X gate is applied to both qubits to correct $q_d$ and reset $q_M$. 
    \item \texttt{Z\_c0}/\texttt{Z\_c1}: The measurement qubit ($q_M$) is put in a deterministic measurement state: $|0\rangle$ (\texttt{Z\_c0}) or $|1\rangle$ (\texttt{Z\_c1}), the latter by applying an $X$ gate before the measurement. In both cases a $Z$ gate is applied to the data qubit $q_d$ and $X$ to $q_M$ conditioned on the measurement being `1'. In the \texttt{Z\_c1} circuit a $Z$ gate is applied on $q_d$ before the feedforward operation. The end result is an identity on the data qubit via no gate (\texttt{Z\_c0}) or two $Z$ gates (\texttt{Z\_c1}). The measurement qubit is reset by the conditional $X$.
    \item \texttt{I\_c0}/\texttt{I\_c1}: Same as \texttt{Z\_c0}/\texttt{Z\_c1} except all $Z$ gates are replaced by identity gates. This circuit performs a feedforward operation whose impact on the data qubit is independent of the measurement outcome, so assignment error does not induce an error on the data qubit.
    \item \texttt{Delay}: A delay time on data and measurement qubits. If the delay time is set to the measurement time and the feedforward time, this gives a baseline on the operational error due to idling errors during the measurement/feedforward operations. Therefore, this circuit is a baseline control. 
\end{itemize}

\section{Relevant Errors and Simulations \label{sect:errors}}

There are a number of potential error terms that may arise in dynamic circuits, in addition to standard gate errors ($\epsilon_G$), for example:
\subparagraph{Idling Errors:}
\begin{itemize}
    \item $T_1/T_2$  Error ($\epsilon_{\tau})$: Measurements and feedforward take time $\tau$ to complete and finite $T_1/T_2$  times result in an incoherent error on the data qubits.
    \item Detuning Error: Coherent Z-phase accumulating during the idle time, due to drifts of the qubits' frequencies and charge-parity fluctuations~\cite{shirizly2024dissipative}. 
    \item ZZ: Coherent ZZ crosstalk between the measurement and data qubit, when they are coupled together.
\end{itemize}
\subparagraph{Readout/Measurement Errors:}
\begin{itemize}
    \item Readout Assignment Error ($\epsilon_R$): This error is the probability that the readout assigns the wrong classical bit to the output register, e.g., due to measurement noise. This is not typically symmetric, i.e., the probability of measuring `1' when the state is $\ket{0}$ is different than the probability of measuring `0' when the state is $\ket{1}$. This is not an error that occurs in standard RB.
    \item QND Error ($\epsilon_{QND}$): Probability that the measurement ends in a different state than it started; may also induce assignment error if the state-flip influences the measurement outcome.
    \item Measurement-Induced Z-Phase and Collisions: Discussed in more detail in Ref.~\cite{Govia_2023,chen:2022}. Z-phase errors can be strongly mitigated by dynamical decoupling.  
\end{itemize}
The focus of this work is assignment error ($\epsilon_R$), since that is not well-characterized in other protocols. In particular, $\epsilon_R$ is not cumulative if the operation is just repeated on the same qubit, as opposed to gate errors which are cumulative in standard RB. By using feedforward we can make assignment error cumulative on the data qubits. In our simulations, we assume the feedforward operation itself is perfect, but that it adds finite time to the circuit~\cite{vazquez:2024}, which is part of the $T_1/T_2$  error term.  To get an approximate idea for the ordering of error terms, we show typical errors for the qubits used in this work on a 127 qubit fixed-frequency `Eagle' processor in Table~\ref{tab:err} (see \apporsm{sec:device}). We can see from this table that assignment error and measurement time are dominating factors.\\

The key circuits we identified in the previous section, and that will be explored in this work, have cascading complexity of errors. The \texttt{Delay} circuits are only sensitive to the idling errors. The \texttt{I\_c0}/\texttt{I\_c1} probe the same timescales as the \texttt{Delay} circuits, but by virtue of including a \emph{real} measurement and feedforward, they test not only for idling errors during the true timescales of these operations, but also for measurement-induced Z-phase and/or collisions. The feedforward introduced here does not map assignment error to data-qubit error since the feedforward operation is the identity, but it does check for any extreme feedforward error, e.g., the feedforward accidentally runs the wrong gate entirely (something we assume is extremely unlikely, but \emph{could} happen). The measurement qubit can also be observed to check for ``out-of-model'' leakage or large assignment/QND error, though only in a qualitative sense. The \texttt{Z\_c0}/\texttt{Z\_c1} circuits add the complexity of a feedforward operation that maps assignment error to the data qubit. Finally \texttt{H\_CNOT} probes all of these elements in addition to two-qubit gate error. \texttt{H\_CNOT} can be seen as similar to a primitive for the checks in an error correction code, but connecting the error rates to code thresholds is beyond the scope of this work. 

\subsection{Theory}

Given the error terms above, it is difficult to come up with general theory expressions for completely general error maps, particularly given the complexity of the measurement operations. However, we can derive lowest order expressions for the assignment error, which is of particular interest given that it is the novel aspect of this protocol. The main results of this section are calculations of the data-qubit average gate error $\epsilon_{\mathcal{F}}$ due to measurement on the measured qubit, given in the following \eq{eqn:Zc0Zc1} for \texttt{Z\_c0}/\texttt{Z\_c1} circuits and \eq{eqn:HCNOT} for the \texttt{H\_CNOT} circuit.
\begin{align}
    &\varepsilon_{\texttt{Z\_c0}/\texttt{Z\_c1}} = \frac{4}{9}\epsilon_R + \mathcal{O}\left(\epsilon_R^2\right), \label{eqn:Zc0Zc1}\\
    &\varepsilon_{\texttt{H\_CNOT}} = \frac{2}{3}\epsilon_R. \label{eqn:HCNOT}
\end{align}
For simplicity, we have assumed that the assignment error is symmetric, i.e.~independent of the state of the measured qubit, but asymmetric assignment error would not change the qualitative conclusions of these results. We now expand upon the details of these derivations.\\

First we look at the \texttt{Z\_c0}/\texttt{Z\_c1} circuits. Resetting the measured qubit after each MCM is useful to avoid a temporally correlated data-qubit error induced by decay of the measured qubit, as discussed in Ref.~\cite{jurcevic:2022}. This non-Markovian error on the data qubits typically leads to an abrupt ``divot'' in their survival probability decay curve as the error model switches mid-circuit \cite{Govia_2023}. Unfortunately, as we now describe, even with reset, our protocols are not immune to non-Markovian data-qubit error induced by a Markovian error channel on the full system. However, as described below, we believe the form of this non-Markovianity is more benign.\\

As we will now explain, a single assignment error will induce two correlated data-qubit errors. Thus, for the data qubits a MCM assignment error looks non-Markovian. However, in our simulations we have found that the data-qubit survival probability still decays approximately exponentially in the presence of MCM assignment error. We explain this minimal non-Markovian deviation by the fact that i) the correlated error only persists for two adjacent MCMs, and ii) the induced data-qubit error is maximally depolarizing but occurs with low probability (given by the assignment error). The second fact in particular leads to the most significant modification to the decay curve occurring at short depths.\\

Our error model for the \texttt{Z\_c1} protocol with assignment error is as follows (the \texttt{Z\_c0} protocol follows analogously). We consider the action of the entire circuit, starting with the $X$-gate on the measured qubit and $Z$-gate on the data qubit, and ending with the conditional operation. If there is no assignment error, then regardless of initial state, the measured qubit is mapped to $\ket{0}$. If the measured qubit started in $\ket{0}$, the net effect on the data qubit of the unconditional $Z$-gate and the conditional operation is identity, while if the measured qubit started in $\ket{1}$, the net effect is a $Z$-gate. If an assignment error occurs, then the situation is reversed: the measured qubit is always mapped to $\ket{1}$, and a net identity (Z-gate) occurs on the data qubit if the measured qubit was initially in $\ket{1}$ ($\ket{0}$). Averaging over the Clifford twirl realizations on the data qubit, the net effect of the $Z$-gate is a depolarizing channel. \\

This error model can be described by the transfer matrix
\begin{align}
    \nonumber\mathcal{T} &= \left(1-\epsilon_R\right)\left(\superketbra{0}{0}\otimes\mathcal{I} + \superketbra{0}{1} \otimes \mathcal{D}_{-\frac{1}{3}}\right) \\
    & + \epsilon_R\left(\superketbra{1}{0}\otimes \mathcal{D}_{-\frac{1}{3}} + \superketbra{1}{1}\otimes\mathcal{I} \right),
\end{align}
where $\superket{j}$ is a vectorization of the density matrix $\ketbra{j}{j}$, and the action of the superoperator $\superketbra{i}{j}$ is the project and prepare channel $\rho\rightarrow \bra{j}\rho\ket{j}\ketbra{i}$. Calligraphic letters indicate quantum channels and their representation as superoperators that act on vectorized density matrices, such as $\mathcal{I}$ the identity map and $\mathcal{D}_q$ the single-qubit depolarizing channel with depolarizing parameter $q$, defined by
\begin{align}
    \mathcal{D}_q(X) = qX + \frac{1-q}{2}{\rm Tr}\left[X\right]\identity.
\end{align}
Note that the parameter $q$ does not correspond to a probability, and for an $n$-qubit depolarizing channel can take negative values as large (in absolute value) as $1/(1-4^n)$. The limit of $q=-1/3$ for a single qubit occurs when Clifford twirling an error channel that is a Pauli gate, as is the case for the \texttt{Z\_c0}/\texttt{Z\_c1} protocols where assignment error induces a Z-Pauli error on the data qubit.\\

At depth $d$ the data-qubit survival probability is given by
\begin{align}
    P(0) = \superbra{\identity}\otimes\superbra{0}\mathcal{T}^d\superket{0}\otimes\superket{0},
\end{align}
where we have assumed we start in $\ket{0}$ for both qubits, and the action of $\superbra{\identity}$ is to trace-out the measured qubit. While the transfer matrix describes a Markovian process for the total system, from the perspective of the data qubit alone it is non-Markovian, as its evolution depends on the history of the measured qubit. Mathematically, this is captured by the fact that for $d> 1$ 
\begin{align}
   P(0) = \superbra{\identity}\otimes\superbra{0}\mathcal{T}^d\superket{0}\otimes\superket{0} \neq \superbra{0}\mathcal{T}_{\rm eff}^d\superket{0},
\end{align}
for any transfer matrix $\mathcal{T}_{\rm eff}$ acting on the data qubit alone that is a valid completely-positive and trace-preserving quantum channel. This is easily verified by the fact that
\begin{align}
\left(\superbra{\identity}\otimes\mathcal{I}\right)\mathcal{T}^2\left(\superket{0}\otimes\mathcal{I}\right) \neq  \left[\left(\superbra{\identity}\otimes\mathcal{I}\right)\mathcal{T}\left(\superket{0}\otimes\mathcal{I}\right)\right]^2.
\end{align}
As the data-qubit evolution is non-Markovian, the qubit's survival probability will not decay exponentially. However, to leading order we find approximate exponential decay with
\begin{align}
    \alpha_{\texttt{Z\_c0}/\texttt{Z\_c1}} = 1-\frac{8}{9}\epsilon_R + \mathcal{O}\left(\epsilon_R^2\right),
\end{align}
from which we obtain the expression for average gate error in \eq{eqn:Zc0Zc1}. \\

Evaluation of the transfer matrix for the \texttt{H\_CNOT} protocol in the presence of assignment error is more straightforward. For the measured qubit, the action of the measurement and controlled operation is a qubit reset. If the assignment is correct, the measured qubit is mapped to $\ket{0}$, and if the assignment is incorrect, then the measured qubit is mapped to $\ket{1}$. For the data qubit, the combined action of the CNOT and controlled operation is a net identity with no assignment error, and results in a bit-flip when assignment error occurs. As before, Clifford twirling this bit-flip results in a depolarizing channel.

The transfer matrix for this error model is given by
\begin{align}
    \mathcal{C} = \left(1-\epsilon_R\right)\superketbra{0}{\identity}\otimes\mathcal{I} + \epsilon_R\superketbra{1}{\identity} \otimes \mathcal{D}_{-\frac{1}{3}},
\end{align}
where the action of the channel $\superketbra{j}{\identity}$ is to map any input state to the output state $\ketbra{j}$. Unlike for the \texttt{Z\_c0}/\texttt{Z\_c1} protocol, the evolution for the \texttt{H\_CNOT} is Markovian for the data qubit alone, and is in fact a depolarizing channel given by
\begin{align}
   \nonumber \mathcal{C}_{\rm eff} &= \left(\superbra{\identity}\otimes\mathcal{I}\right)\mathcal{C}\left(\superket{\identity}\otimes\mathcal{I}\right) \\&= \left(1-\epsilon_R\right)\mathcal{I} + \epsilon_R\mathcal{D}_{-\frac{1}{3}} = \mathcal{D}_{\left(1-\frac{4}{3}\epsilon_R\right)}.
\end{align}
As the data-qubit channel is depolarizing, we immediately get that the survival probability decays exponentially with an average gate error given by Eq.~\eqref{eqn:HCNOT}. 

\subsection{Simulation} \label{sec:simu}

\begin{figure}
\centering
\includegraphics[width=0.45\textwidth]{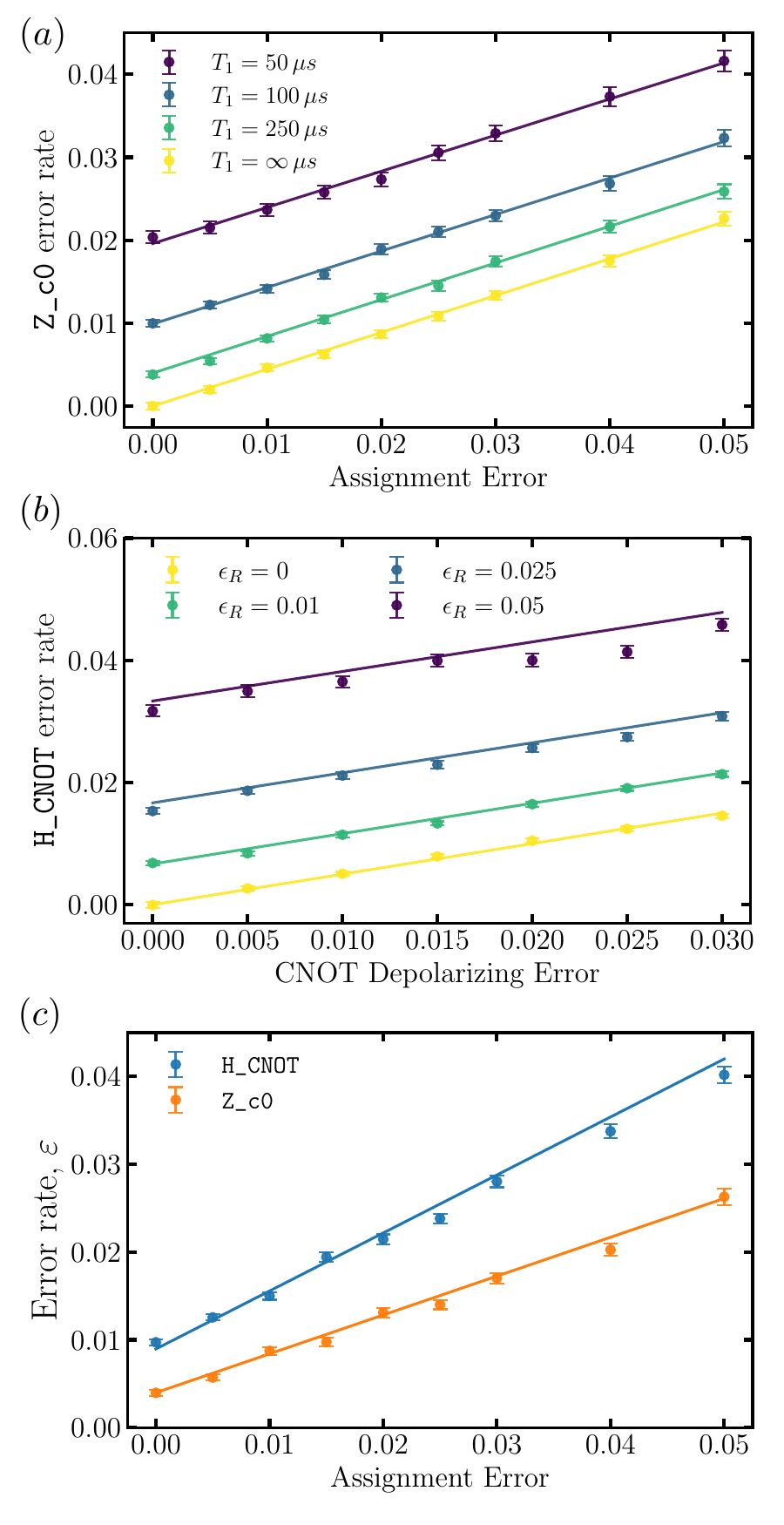}
\caption{(a) Simulation results of \texttt{Z\_c0} for varying values of $T_1$ for the data qubit ($T_1=T_2$) as a function of readout assignment error $\epsilon_R$. We fit to the first-order expression as described in the main text. (b) Simulation results of \texttt{H\_CNOT} for varying values of $\epsilon_R$, as a function of the CNOT depolarizing error. (c) Simulation results of \texttt{H\_CNOT} compared to \texttt{Z\_c0} at $T_1=T_2=250\,\mu{\rm s}$. The \texttt{H\_CNOT} block has an added 2Q depolarization probability of 0.01 ($7.5\times10^{-3}$ gate error). Both fit to the first order expression, with 2/3 and 4/9 linear dependence on the assignment error for \texttt{H\_CNOT} and \texttt{Z\_c0} respectively.
\label{fig:t1_readout}}
\end{figure}

To understand the protocol under more general noise  we have to turn to numerical simulations. We build a noise model in the Qiskit Aer simulator~\cite{aer} and run the same procedure as we use in \S~\ref{sect:exp} for running the experiment, i.e., simulate the RB sequences and fit to an exponential decay. As we focus here on low error rates, and for accurate comparison to the theory, we extract the error per block as in interleaved RB, $\varepsilon_{\mathcal{F}_i} = \frac{1}{2}\left(1-\alpha_{\mathcal{F}_i}/\alpha\right)$, with $\alpha$ the decay rate of a standard 1Q RB reference curve. We use the 1Q gate error on the data qubit to define $\alpha$.\\

In all simulations, we use a 1Q gate depolarizing error of $5\times10^{-4}$ ($2.5\times10^{-4}$ gate error rate). The remaining noise terms are written in each figure and the idle noise terms assume a total duration of the dynamic circuit block of $\tau=2\,\mu{\rm s}$. In Fig.~\ref{fig:t1_readout}(a) we show the data-qubit error rate of the \texttt{Z\_c0} block with symmetric readout assignment error of the measured qubit, and amplitude/phase damping of the data qubit with $T_1=T_2$. The error rate depends linearly on the assignment error in this parameter range for fixed amplitude and phase damping. This fits with nice agreement to the combined error rate of the individual terms 
\be  \varepsilon_{\texttt{Z\_c0}} = 1 - \left(1-\frac{4}{9}\epsilon_R\right)\left(1-\epsilon_{\tau}\right), \ee
where the readout assignment prefactor is given in \eq{eqn:Zc0Zc1}, and the idling error is given by 
\be \epsilon_{\tau} = \frac{2}{3}\left(\frac{3}{4} - \frac{e^{-\tau/T_1}}{4} - \frac{e^{-\tau/T_2}}{2}\right). \ee
For this model, with symmetric error and no decay of the measured qubit, the results of \texttt{Z\_c0} and \texttt{Z\_c1} are the same. 
The error rate of the \texttt{H\_CNOT} block is shown in \fig{fig:t1_readout}(b), with linear dependence on the CNOT error. This fits well with 
 \be \varepsilon_{\texttt{H\_CNOT}} = 1 - \left(1-\frac{2}{3}\epsilon_R\right)\left(1-\frac{2}{3}\epsilon_{2Q}\right), \ee with $\epsilon_{2Q}$ the CNOT gate error, and $\epsilon_R$ the readout assignment error from \eq{eqn:HCNOT}.  In order to highlight the different linear dependence on the assignment error we show a comparison of \texttt{Z\_c0} to \texttt{H\_CNOT} in Fig.~\ref{fig:t1_readout}(c). Here there is also finite $T_1/T_2$,  thus the error of \texttt{H\_CNOT} is approximated by \be\label{eq:approxHCNOT}  \varepsilon_{\texttt{H\_CNOT}} = 1 - \left(1-\frac{2}{3}\epsilon_R\right)\left(1-\epsilon_{\tau}\right)\left(1-\frac{2}{3}\epsilon_{2Q}\right). \ee \\

\begin{figure}
\centering
\includegraphics[width=0.48\textwidth]{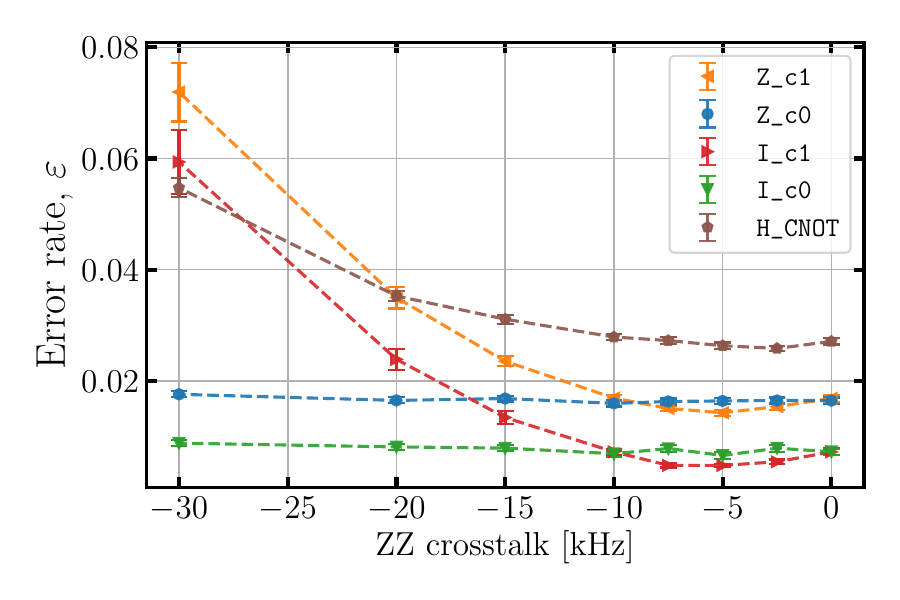}
\caption{Simulation results for 5 of our dynamic circuit blocks as a function of ZZ crosstalk between the measured and the data qubit. Noise parameters are taken to be $\epsilon_R=0.02$, $T_1=T_2=250\,\mu{\rm s}$, $\epsilon_{2Q}=0.01$ and $\Delta=10$ kHz. 
\label{fig:zz}}
\end{figure}

During the finite-time dynamic circuit blocks, there are also coherent noise terms. As an example, we focus on ZZ crosstalk and single qubit detuning, which are modeled with the Hamiltonian, 
\begin{equation}
\mathcal{H}/\hbar =\sum_{i}\frac{1}{2}{\Delta}_{i} (I - \sigma^z_{i}) + \sum_{\left\langle i,j\right\rangle }\frac{1}{2}{\zeta}_{ij} (I - \sigma^z_{i}) ( I - \sigma^z_{j}),
\end{equation}
where $\Delta_i$ is the detuning of qubit $i$ and $\zeta_{ij}$ is the crosstalk between qubits $i,j$. In \fig{fig:zz} we show simulation results of our dynamic circuit blocks as a function of the crosstalk $\zeta$, with fixed $\epsilon_R=0.02$, $T_1=T_2=250\,\mu{\rm s}$, $\epsilon_{2Q}=0.01$ and $\Delta=10$ kHz. The larger (absolute) $\zeta$ in \fig{fig:zz} are typical values for connected qubits on `Eagle' processors. However, this is significantly smaller on tunable-coupling `Heron' processors \cite{nowik2024modeling,mckay:2023}. The low crosstalk limit can also represent disconnected qubits. As expected, the blocks where the measured qubit is prepared in $\ket{0}$ are not sensitive to such crosstalk. \texttt{Z\_c1} and \texttt{I\_c1} are affected similarly with higher error from larger crosstalk, except when $\zeta \approx -\Delta/2$, as here the crosstalk and the detuning cancel each other, causing these blocks to have lower error than \texttt{Z\_c0} and \texttt{I\_c0}. The \texttt{H\_CNOT} block includes additional gate error and also is more sensitive to the readout assignment, as noted above, thus at low crosstalk it has higher error. However due to the measured qubit state being prepared in $\ket{+}$, which is then collapsed with equal probability to $\ket{0}$ or $\ket{1}$, it is less sensitive to $\zeta$ than \texttt{Z\_c1/I\_c1}. Since we do not include measurement-induced phase shift and collisions in this simulation, the \texttt{Delay} block is trivially the same as \texttt{I\_c0}, and is not shown here.  

\section{Experimental Results \label{sect:exp}}

In the following section we experimentally measure the different dynamic circuit blocks $\mathcal{F}$ that were outlined in \S~\ref{sect:protocol}. We run these experiments on a 127 qubit fixed-frequency, fixed-coupler IBM Quantum Eagle processor `ibm\_pinguino1'. The specification of the device and the qubits used in this work are given in \apporsm{sec:device}, and the median values of device parameters are given in Table~\ref{tab:err}. \\

\begin{figure}
\centering
\includegraphics[width=0.47\textwidth]{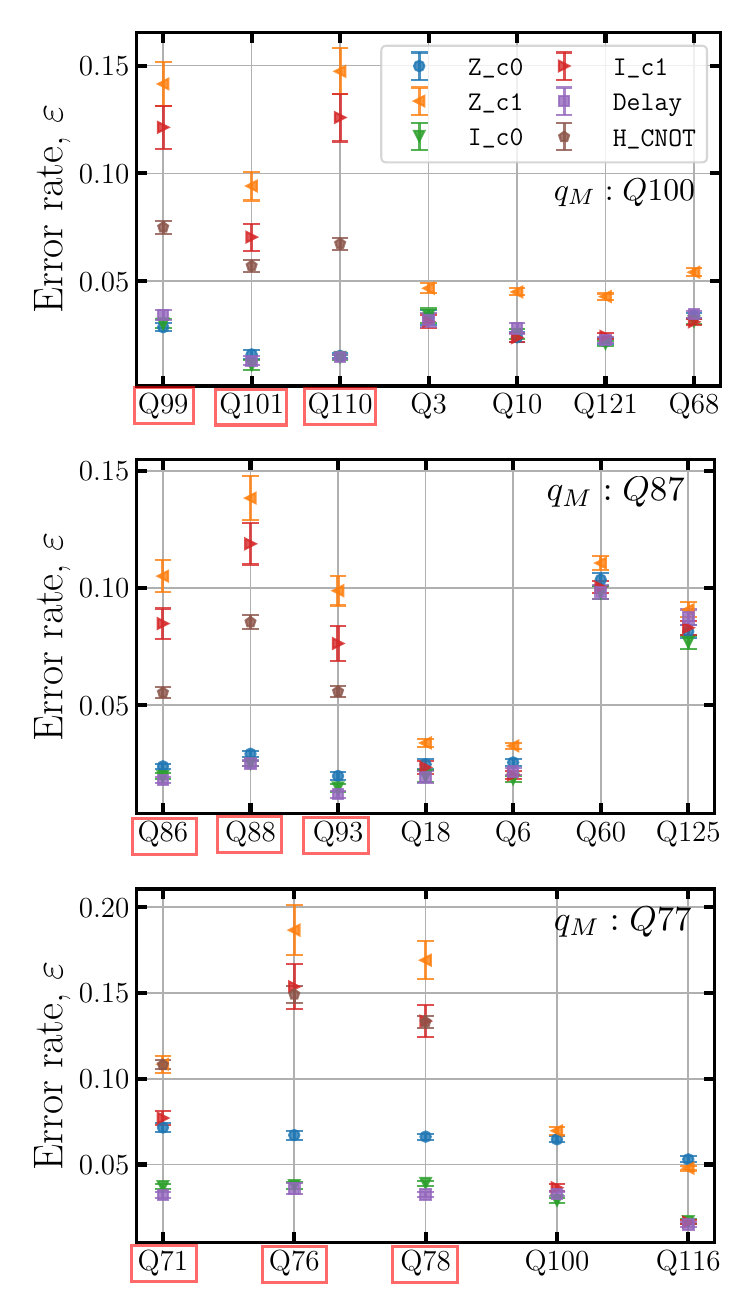}
\caption{
The error rate for the different protocols as measured on three sets of qubits. The highlighted qubits are the nearest-neighbor (connected) qubits and the rest have large distance separation on the device. The large difference between these groups can be ascribed to crosstalk. The ratio between \texttt{Z\_c0} and \texttt{Z\_c1} is due to asymmetric behavior of the measured qubits, see text for details. In the case of the set with measured qubit 77, the assignment error is high, $\approx0.1$. No dynamical decoupling is performed in this experiment. 
\label{fig:full_protocol}
}
\end{figure}

We probe the three qubit sets listed in Fig.~\ref{fig:layout} with all the different protocols: \texttt{Z\_c0}/\texttt{Z\_c1}, \texttt{I\_c0}/\texttt{I\_c1}, \texttt{H\_CNOT} and \texttt{Delay} (control). For each set there are 3 nearest neighbor data qubits and additional separated data qubits. In the data shown in \fig{fig:full_protocol} we use 20 random Clifford sequences with 300 shots each, and 5 Clifford gates per dynamic circuit block $\mathcal{F}$.  The experiments are performed separately for each set, with distinct (uncorrelated) random Clifford sequences on the data qubits. We can only perform the \texttt{H\_CNOT} circuit block on the nearest neighbor qubits, but the other circuit blocks can be performed regardless of connectivity. \\

The results are significantly different whether the data and measurement qubit are nearest neighbors (connected) or not (disconnected). For the disconnected qubits (qubits that are not highlighted by a red box in Fig.~\ref{fig:full_protocol}), we observe similar results for \texttt{I\_c0}/\texttt{I\_c1} and \texttt{Delay}, which indicates that there is no long range crosstalk or measurement collision. This is unsurprising as the qubits are not connected. However, as \texttt{Z\_c0}/\texttt{Z\_c1} maps the measurement qubit assignment error to the data qubits we observe higher error for these protocols. There is also a notable difference between \texttt{Z\_c0} and \texttt{Z\_c1}, which has a variety of possible causes. Firstly, there is a different assignment error whether the measured qubit is in $\ket{0}$ or $\ket{1}$, partly due to $T_1$ decay during the measurement \cite{Thorbeck24} impacting the assigned bit value, but also be due to other issues such as leakage \cite{chen:2022,Sundaresan:2023}. Secondly, for \texttt{Z\_c1} a $T_1$ decay of the measurement qubit that is not detected by the current measurement (QND error without assignment error) will still impact subsequent parts of the circuit. Finally, there is further $T_1$ decay of the measurement qubit during the feedforward delay time. We show the complete raw data traces in \apporsm{sect:rawdata}. In each experiment we also have access to the measurement, which can show evidence of QND error and measurement leakage, which we discuss in \apporsm{sect:meas}. \\

When the qubits are nearest neighbor (qubits that are highlighted by a red box in Fig.~\ref{fig:full_protocol}), we see much larger error rates which is indicative of measurement or idling crosstalk. Comparing \texttt{I\_c0} and \texttt{Delay}, we can see evidence for measurement-induced Z-phase or crosstalk, mainly on the set with measured qubit 77. The effect of static ZZ-crosstalk (independent of the measurement) is also shown in the higher error of the \texttt{I\_c1} block. As was the case for the disconnected qubits, the mapping of assignment error of the measured qubit leads to higher error for the \texttt{Z\_c0/Z\_c1} blocks compared to their identity references \texttt{I\_c0/I\_c1}. The \texttt{H\_CNOT} block involves a combination of all of these errors: we expect an averaged ZZ-crosstalk error due to the measured qubit preparation in the $\ket{+}$ state, together with the assignment error mapping to the data qubits, which as previously discussed scales differently compared to the \texttt{Z\_c0/Z\_c1}.  \\

\begin{figure*}
\centering
\includegraphics[width=0.91\textwidth]{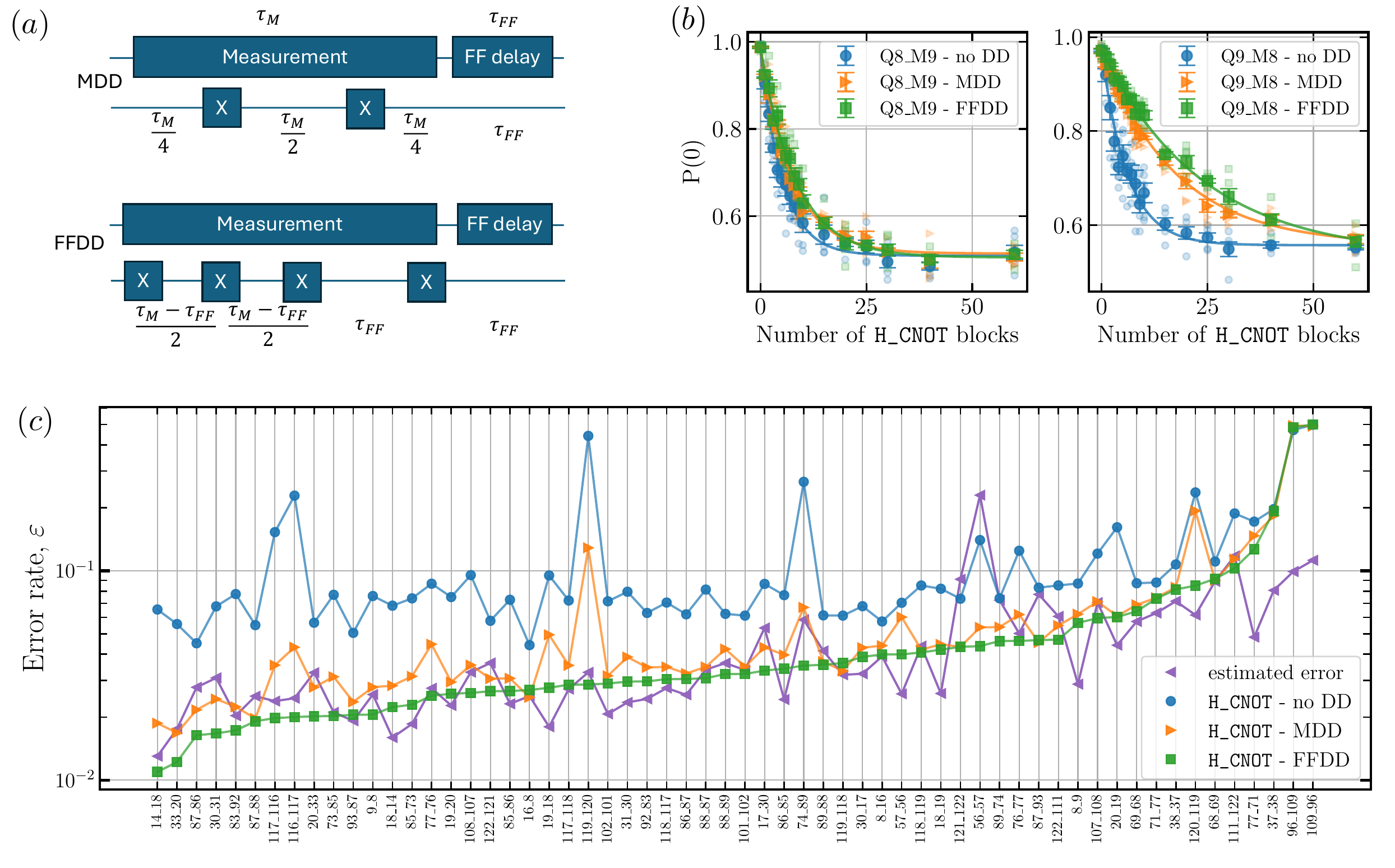}
\caption{\label{fig:CNOT with DD} (a) Dynamical decoupling (DD) sequences that we consider (not to scale). (Top) Measurement DD (MDD), which involves X2 DD within the measurement time on the data qubit. (Bottom) FFDD, to suppress the ff delay time, where the full delay is sliced to two sequences of X2, one with the ff delay time and the second with the remaining delay $\tau_{M}-\tau_{FF}$ assuming $\tau_{M}>\tau_{FF}$. For the device used in this work, $\tau_M=1512~\text{ns}$ and $\tau_{FF}=1060~\text{ns}$. (b) Experimental results of the \texttt{H\_CNOT} protocol with and without DD on qubits 8 and 9. (c) Error rate of 30 connected pairs, extracted as in (b), each pair measured separately. The FFDD outperforms the MDD in almost all pairs, although the biggest improvement is gained by doing any type of DD. Once DD is applied the data agrees well to a model adding the assignment error and the CNOT error as reported from the device, see text for details.
}
\end{figure*}

Since the dynamic circuit blocks involve relatively long idle times and we see clear evidence of crosstalk, this presents an opportunity to deploy dynamical decoupling (DD). However, there are some limitations on DD that can be applied due to the feedforward operations. The measurement duration is part of the scheduled circuit, and so a DD sequence can be applied there. For feedforward operations there is an additional time $\tau_{\text{FF}}$ which is a complicated function of the various classical electronic components involved and the complexity of the classical feedforward conditional statement. These times are not reported, but can be determined experimentally by Ramsey experiments, or, as discussed next, by optimizing the DD signal. Unfortunately, a gate sequence cannot be applied during the feedforward latency time. Therefore, we explore here two cases of possible DD, as shown schematically in \fig{fig:CNOT with DD}(a). The first one (MDD), applies an $X-X$ sequence during the measurement time $\tau_M$, and the second (FFDD), is a DD sequence that utilizes $\tau_{\text{FF}}$ as one arm of a DD sequence (similar to the sequence demonstrated in Ref.~\cite{baumer2024quantum}). FFDD is only possible if $\tau_{\text{FF}}$ is known, which must be measured or estimated independently; in the data we show here $\tau_{\text{FF}}=1060~\text{ns}$. Focusing only on the data qubit, these DD sequences will suppress ZZ-crosstalk with the measured qubit. However for suppressing crosstalk on adjacent qubits as well, staggered DD sequences are needed~\cite{zhou2023quantum,shirizly2024dissipative,seif2024suppressing}, which can be applied together with the FF time limitations by increasing the total delay~\cite{vazquez:2024}. We do not explore these here. \\

In Fig.~\ref{fig:CNOT with DD}(b) we show an example of the decay curves of the \texttt{H\_CNOT} dynamic circuit block with or without DD, for qubits 8 and 9 playing both the role of measurement and data qubit. Here, we use 5 random Clifford sequences for each sequence length. Fitting to an exponential, the decay of data qubit 9 (8) gives the error rates 0.076 (0.087), 0.028 (0.062) and 0.021 (0.056) for the block with no DD, MDD and FFDD respectively. Without DD the error rates are similar despite a strong asymmetry in the assignment fidelities (0.01 and 0.04 for qubit 8 and 9 respectively). This indicates that coherent noise processes dominate for both qubits. The use of DD removes this error, and then the error rates are very asymmetric due to the asymmetric assignment fidelities. Such asymmetries motivate benchmarking both directions of each pair for choosing which qubits to use when utilizing dynamic circuits. \\

We similarly probe 29 additional connected qubit-pairs in both directions, and their error rates are shown in \fig{fig:CNOT with DD}(c). In all cases the data is much improved through the use of DD, where the median errors are 0.077, 0.042 and 0.032 for no DD, MDD and FFDD respectively. When the coherent dynamics are mostly suppressed, we can approximate the expected error rate as in \eq{eq:approxHCNOT} with the individual parameters of each qubit. This estimated error and the experimental FFDD data agree qualitatively, which indicates that the dominant errors in the dynamic circuit block are due to the dissipation during idling, residual CNOT error, and assignment error.

\section{Discussion and outlook}

In the preceding sections, we have introduced a protocol to interleave dynamic circuit blocks $\mathcal{F}$ (which are net-identity operations) on a measurement qubit into 1Q RB sequences on data qubits. In doing so, we can probe several important error terms for dynamic circuits, and importantly, this protocol can inject readout assignment errors into the data-qubit RB sequence. As a result, readout assignment error accumulates on the data qubit, and this can be used to measure small readout errors, as will be necessary in future applications. However, in current generation devices we find that there are still large errors that dominate, particularly decoherence during the long timescale of the measurement and feedforward operations. This is clearly seen by the comparable error rates when interleaving only a delay operation. Furthermore, we observed the importance of utilizing dynamic decoupling during these dynamic circuit blocks to suppress the impact of coherent error. \\

Because of the particular nature of measurement, this protocol does not return a result that could be viewed in the same lens as the average gate fidelity. Furthermore, there is additional work required to relate the error rates measured here to the error rates of an error correction code and to other methodologies for holistically benchmarking codes~\cite{gicev:2024,liepelt:2024,hockings:2024}. For example, the \texttt{H\_CNOT} block is similar in structure to an error correction stabilizer measurement, but in an actual quantum error correction code this would involve $M$ data qubits to one measurement qubit, which is not the circuit we use here. However, the error reported by the \texttt{H\_CNOT} protocol measured separately for all $M$ data qubits may be useful as an estimated upper bound to the error rate of the full stabilizer. Such questions are left to future research. In the interim, the protocol defined here has tremendous diagnostic value and is a fast and simple method for detecting faults in dynamic circuits. Furthermore, running our protocols in parallel across a device should reveal dynamic circuit crosstalk, which is an extension worthy of further consideration.

\acknowledgements

We acknowledge valuable conversations with Diego Rist\`e, Maika Takita, Seth Merkel and Haggai Landa. We thank Maika Takita and Doug McClure for their feedback on our manuscript. Research was sponsored by the Army Research Office and was accomplished under Grant Number W911NF-21-1-0002. The views and conclusions contained in this document are those of the authors and should not be interpreted as representing the official policies, either expressed or implied, of the Army Research Office or the U.S. Government. The U.S. Government is authorized to reproduce and distribute reprints for Government purposes notwithstanding any copyright notation herein.

\bibliography{main}

\appendix

\begin{figure}[h!]
\centering
\includegraphics[width=0.48\textwidth]{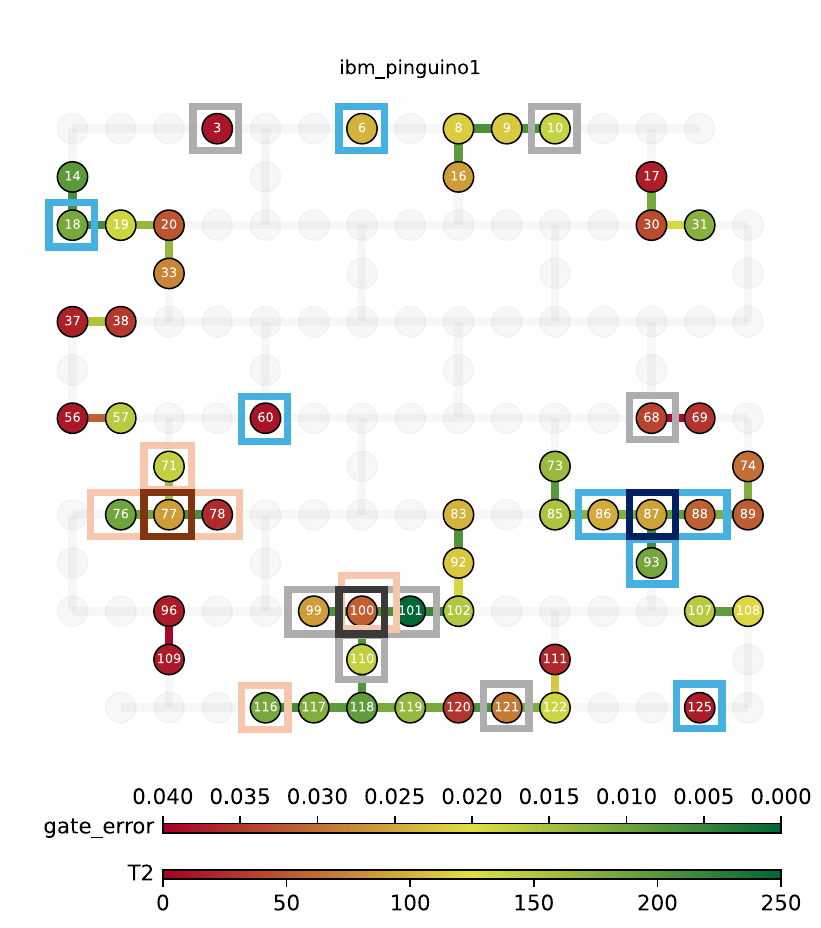}
\centering
\caption{\label{fig:layout} Partial layout of the device considered here (data as of August 4, 2024 which is the date of the experiments), including only the qubits considered in this work. We highlight the three sets used in \fig{fig:full_protocol}: set one is measurement qubit 100, data qubits \{99,101,110,3,10,121,68\}, set two is measurement qubit 87, data qubits \{86,88,93,18,6,60,125\} and set three is measurement qubit 77, data qubits \{71,76,78,100,116\}. We consider neighbors and qubits which are far separated from the measurement qubit. The rest of the plotted qubits are used in \fig{fig:CNOT with DD}.}
\end{figure}

\section{Device Details \label{sec:device}}

The experimental results in this paper are obtained on a 127 qubit fixed-frequency, fixed-coupler IBM Quantum Eagle processor `ibm\_pinguino1'. The layout of this device together with the reported $T_2$ and two-qubit gate errors for the qubits used are shown in \fig{fig:layout}. We give aggregate numbers for errors, coherences and operation times in Table.~\ref{tab:err}. \\

\begin{table}[h!] % put at top of page if possible 
\begin{tabular}{|l|c|}
\hline
$T_1$ & $208\,\mu{\rm s}$ \\ \hline
$T_2$ & $97\,\mu{\rm s}$ \\ \hline
$\tau_{\mathrm{meas}}$ & $1512~\text{ns}$ \\ \hline
$\tau_{\mathrm{2Q}}$ & $660~\text{ns}$ \\ \hline
$\tau_{\mathrm{1Q}}$ & $60~\text{ns}$ \\ \hline
$\epsilon_{R}$ & 2.2e-2 \\ \hline
$\epsilon_{2Q}$ & 9.7e-3 \\ \hline
$\epsilon_{1Q}$ & 2.4e-4 \\
\hline
\end{tabular}
\caption{Median errors ($\epsilon$) and operation lengths ($\tau$) for `ibm\_pinguino1' for the categories listed in \S~\ref{sect:errors}, for the qubits used in this work. We see that measurement assignment error is one of the larger errors and that measurement operations are the longest. While gate times have been trending downward with progress in the field, measurements have been more difficult to shorten in duration. Data pulled from Qiskit on August 4, 2024. \label{tab:err}}
\end{table}

\section{Measurement Data \label{sect:meas}}

\begin{figure}[h]
\centering
\includegraphics[width=0.44\textwidth]{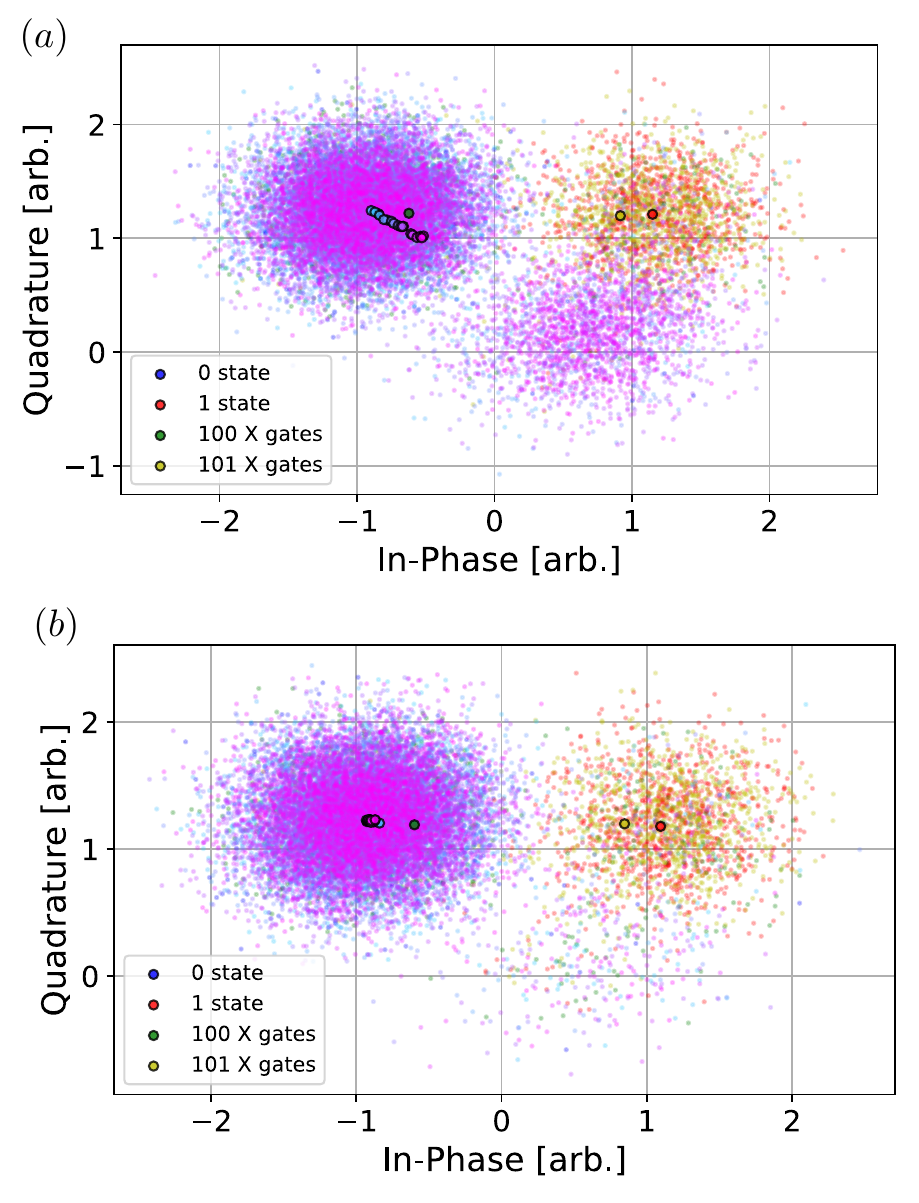}
\caption{Measurement induced leakage. (a) IQ data of repeated X -- Measurement -- Conditional X, similar to the \texttt{Z\_c1} protocol and (b) IQ data of repeated Measurement -- Conditional X as in the \texttt{Z\_c0} protocol. The colors of dynamic circuit block repetitions $N$ are ranging from cyan to purple as the repetitions increase, with $N \in \left\{0, 1, 2, 3, 4, 5, 6, 7, 8, 9, 10, 15, 20, 25, 30, 40, 60\right\}$. The small dots represent each shot and the markers represent the mean values. As the number of repetitions increase, the mean value shifts to higher states when the measured qubit is prepared in $\ket{1}$. Repeated X gates show no leakage by the gates.
\label{fig:leakage}}
\end{figure}

While the protocol focuses mainly on the data qubits, the outcomes of the measurement qubits can also be analyzed. In theory they should quickly latch to a combination of the assignment and QND errors given the reset operation. As seen in Fig.~\ref{fig:all_curves}, some of the measured qubits have notable decay. To further characterize the measurement data, one can measure the full IQ values of the final measurement. An example of such is given in Fig.~\ref{fig:leakage}, where there is notable leakage on these measurements. Such an error is out of the scope of the work here, but important to consider. 

\section{Raw Data \label{sect:rawdata}}

\begin{figure*}
\centering
\includegraphics[width=0.94\textwidth]{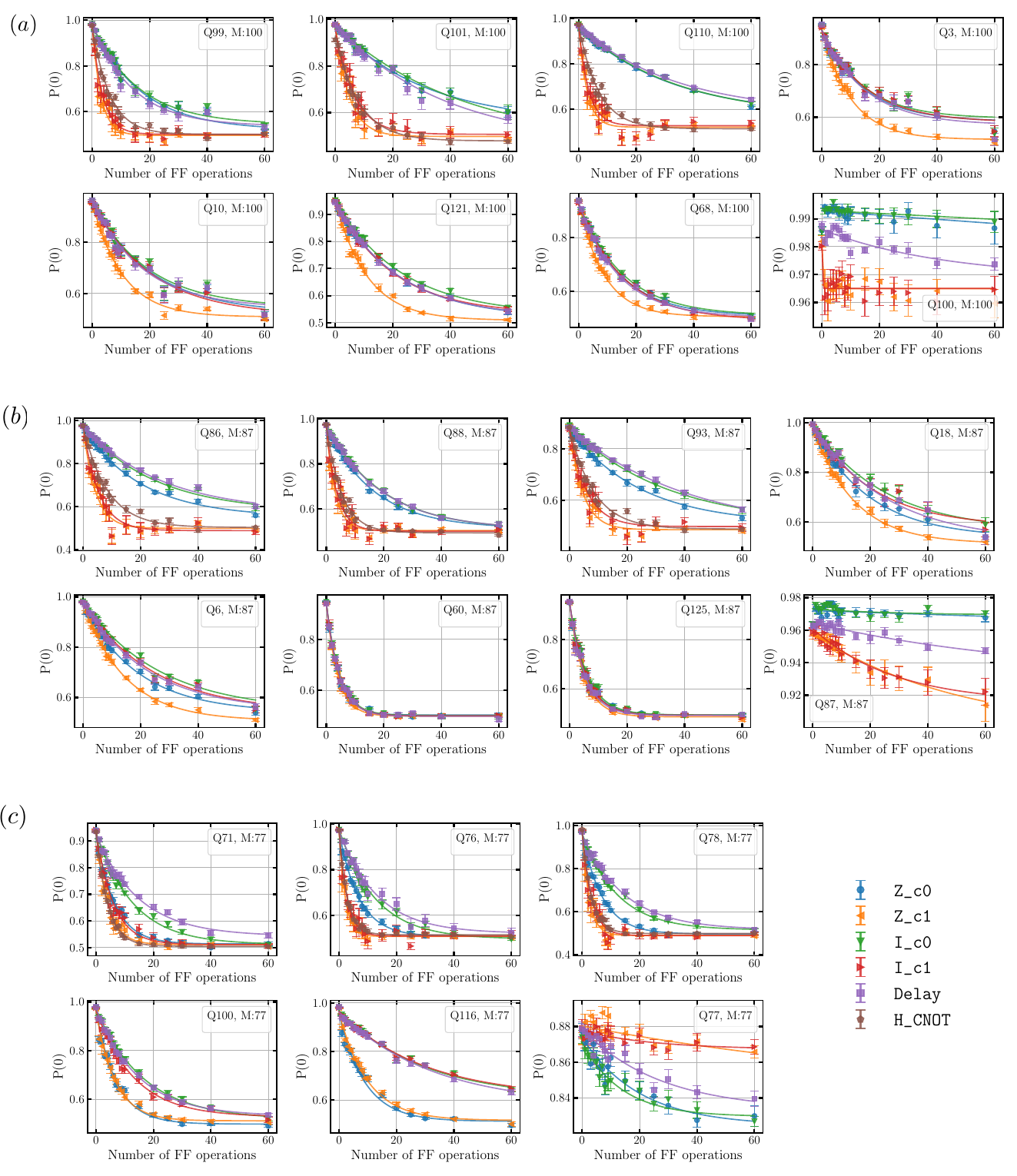}
\caption{\label{fig:all_curves} Individual decay curves of each qubit out of the three sets used in \fig{fig:full_protocol}.
}
\end{figure*}

Here we provide the full curves for the data shown in Fig.~\ref{fig:all_curves}.

\end{document}